\documentclass[twocolumn, pra, showpacs]{revtex4}

\usepackage{graphicx}
\usepackage{amsmath}
\usepackage{subfigure}

\begin{document}

\author{J.~J. Hudson, H.~T. Ashworth, D.~M. Kara, M.~R. Tarbutt, B.~E. Sauer, E.~A. Hinds}
\affiliation{Centre for Cold Matter, The Blackett Laboratory, Imperial College London, London SW7 2AZ, UK}
\title{Pulsed beams as field probes for precision measurement}
\begin{abstract}

We describe a technique for mapping the spatial variation of static
electric, static magnetic, and rf magnetic fields using a pulsed
atomic or molecular beam. The method is demonstrated using a beam
designed to measure the electric dipole moment of the electron. We
present maps of the interaction region, showing sensitivity to (i)
electric field variation of 1.5~V/cm at 3.3\,kV/cm with a spatial resolution of 15~mm; (ii)
magnetic field variation of 5~nT with 25~mm resolution; (iii) radio-frequency
magnetic field amplitude with 15~mm resolution. This new diagnostic technique
is very powerful in the context of high-precision atomic and
molecular physics experiments, where pulsed beams have not hitherto
found widespread application.

\end{abstract}

\maketitle

\section{Introduction}

The structure and interactions of atoms and molecules have been
studied with great precision using atomic and molecular beams
\cite{Scoles, Ramsey}. Some examples include accurate time-keeping
\cite{fountainreview, clockreview}, precise tests of physical
theories \cite{CsParity, haensch1s2s}, measurements of the constancy
of physical constants \cite{haenschconstants, ohye, nh3}, and
searches for physics beyond the Standard Model \cite{Commins94,
EDMPRL, fortson, ICAP}. These kinds of experiments require careful
control and monitoring of stray and applied fields, both electric
and magnetic, throughout the interaction region of the apparatus. In
this paper we develop a novel method for this monitoring, based on
the use of pulsed supersonic molecular beams. Such beams are well
known as useful spectroscopic sources for many applications in
physics and chemistry, mainly because they can provide a wide
variety of molecular species at low temperature and high average
intensity \cite{Scoles}. Here, we point out that the small
spatial/temporal extent of the pulses opens a way to map electric and magnetic fields with high spatial resolution. We
demonstrate this idea using a high-voltage parallel-plate structure
that is also a radio-frequency (rf) transmission-line. This allows
us to drive molecular hyperfine transitions anywhere in the
interaction region. The line centre, lineshape and transition
dynamics then provide information about the electric, magnetic and
rf fields at the place where the transition occurs.

\section{The beam machine}

\begin{figure}[htb]
\begin{center}
\includegraphics[width=7.5cm]{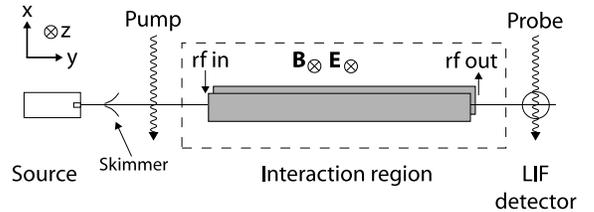}
\caption{Schematic diagram of the beam machine. The molecular packet
issues from the source (in the y-direction) and is skimmed before
being optically pumped into a single hyperfine state. The molecules
enter the magnetically shielded interaction region (dashed line) and
fly through a high-voltage capacitor that doubles as an rf
transmission-line. The electric field points along the z-direction.
The rf-magnetic field is polarised along the x-direction. A small B
field can be applied in the z-direction. After leaving the
interaction region the molecules are detected by laser induced
fluorescence.} \label{machine}
\end{center}
\end{figure}

The apparatus used to demonstrate this method is shown schematically
in Fig.~\ref{machine}. Its primary purpose is to measure the
electric dipole moment of the electron \cite{EDMPRL}, a very
sensitive experiment that requires careful field mapping. A pulse of
YbF molecules is produced in the source every 40~ms by
laser-ablating a solid Yb target into a supersonically expanding gas
jet of Ar and SF$_6$  \cite{TarbuttSource}. The gas pulse passes
through a 2~mm diameter skimmer, placed 60~mm from the source, in
order to form a well-collimated beam. Each pulse of YbF molecules
forms in a small region over a short time (see section
\ref{tofsection}) and cools to give a few times $10^5$ molecules at
the detector in the ro-vibronic ground state. The molecular packets
have a mean velocity of about 590~m/s and a velocity spread of
40~m/s FWHM, corresponding to a translational temperature of a few
Kelvin. The length of a packet anywhere in the interaction region is
therefore approximately $7\%$ of its distance from the source.

\begin{figure}[htb]
\begin{center}
\includegraphics[width=8.5cm]{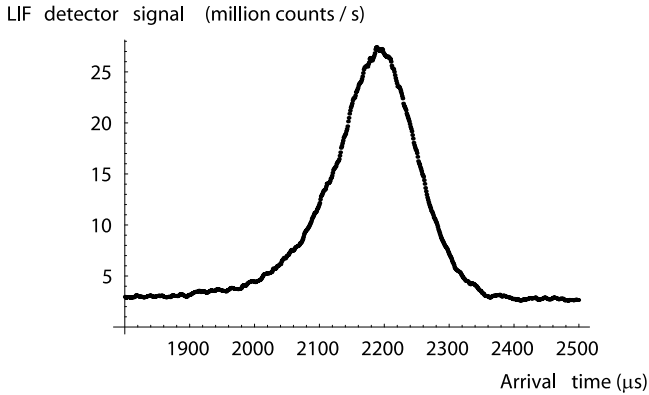}
\caption{A packet of molecules passing through the detector. Each
molecule emits on average 1.9 photons before being lost to a dark
state. Approximately 2\% of the emitted photons are registered. }
\label{tof}
\end{center}
\end{figure}

A single-mode continuous-wave dye laser beam, the pump shown in
Fig.~\ref{machine}, excites the molecules downstream from the
skimmer on the $A^2\Pi_{1/2}-X^2\Sigma^+$ Q(0) transition at 552~nm.
The Doppler width is suppressed by pointing the pump laser beam
perpendicular to the molecular beam. This gives a narrow enough
linewidth to resolve the two X-state hyperfine levels $F=0$ and
$F=1$, separated by 170\,MHz. We tune the pump laser to excite the
$F=1$ population, which becomes depleted as a result. Any remaining
$F=1$ molecules are detected at the end of the 1.3~m-long beam
machine using a second laser beam of the same frequency, the probe
in Fig.~\ref{machine}. The laser-induced fluorescence (LIF) is
monitored by a photomultiplier, a typical pulse (without pump) being
shown in Fig. \ref{tof} versus time of arrival. The 140~$\mu$s width
of this pulse is essentially all due to the velocity spread of the
beam. The temporal resolution of the LIF detection is 7~$\mu$s, and
is due principally to the 4~mm spatial width of the probe laser
beam.

The interaction region, shown in Fig.~\ref{machine}, starts 450~mm
from the skimmer and is 790~mm long. It is magnetically shielded to
reduce the ambient field, and current-carrying wires inside the
shields allow us to apply a magnetic field. Within this region there
is a pair of electric field plates, 750~mm long and 70~mm wide, with
a 12~mm spacing accurate to better than 200~$\mu$m over the full
length. These plates are machined from cast aluminium, then gold
coated to improve the uniformity of the surface potential. The whole
assembly is non-magnetic. With a field across the gap of 15~kV/cm
the leakage current is less than 1~nA.

\begin{figure}[htb]
\begin{center}
\includegraphics[width=7.5cm]{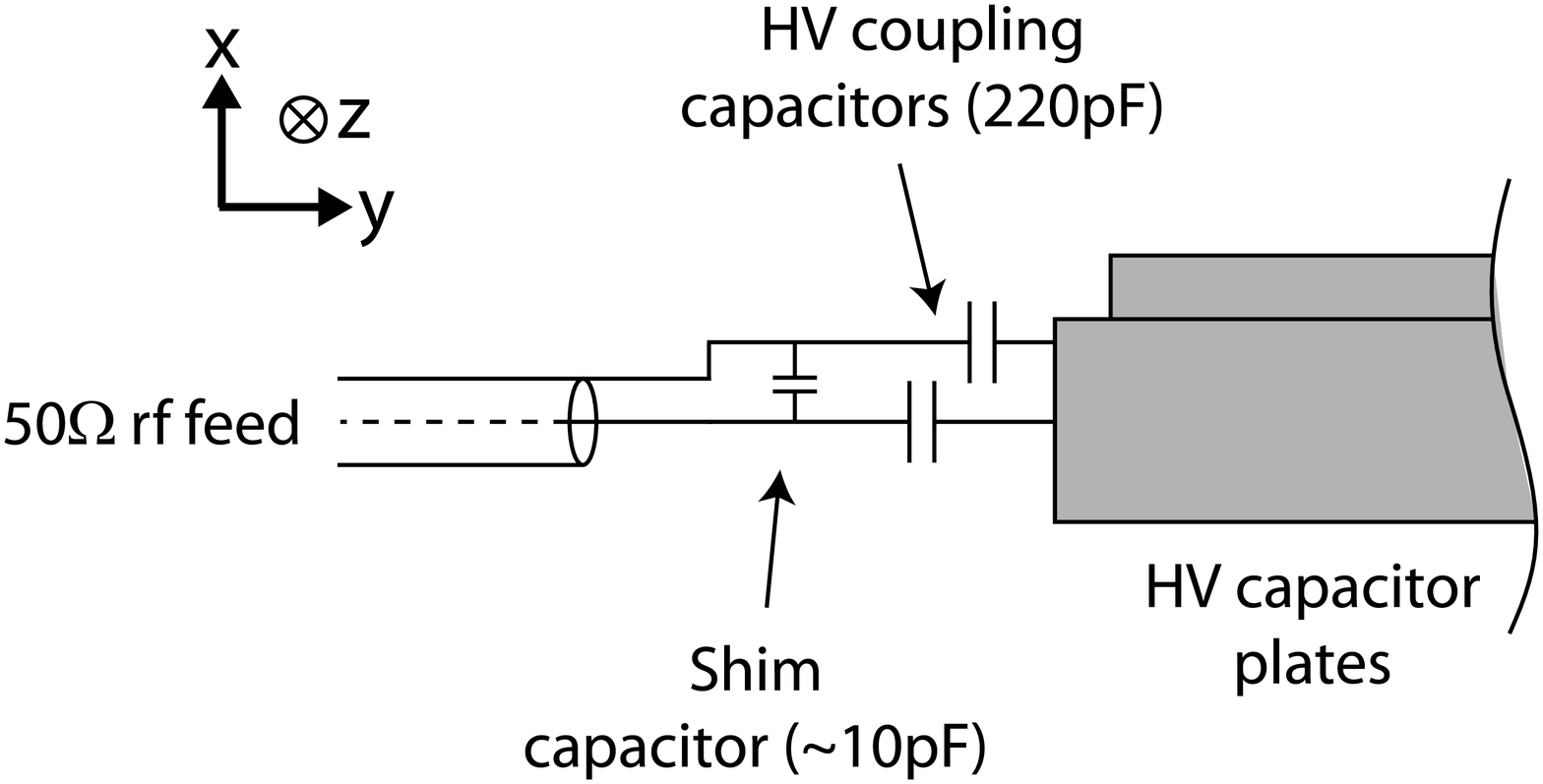}
\caption{rf transmission line coupling. Both ends of the capacitor plates are coupled to 50~$\Omega$ co-axial feeds in this manner. rf is injected in one end of the capacitor and coupled out of the other. The direction can be reversed to check for systematic effects.} \label{coupling}
\end{center}
\end{figure}

The same plate structure also serves as a 34~$\Omega$ transmission
line for the 170~MHz radiation that drives the ground-state
hyperfine transition. This transports the radiofrequency field as a
fundamental TEM wave travelling parallel or antiparallel to the beam
direction. Figure \ref{coupling} shows how the rf is coupled in and
out of the transmission line using high-voltage porcelain 100~pF
capacitors to connect to the 50~$\Omega$ non-magnetic semi-rigid
cable. The connections are shimmed with small ($\sim$10~pF) ceramic
disk capacitors to minimise the reflections at each end. As a packet
of molecules travels along the beamline, we can drive a hyperfine
state changing transition at any desired position within the plate
structure by pulsing the rf field on for a short period of time.

\section{Time of flight and spatial resolution}
\label{tofsection}

It is a key element of our mapping technique that the pulses of
molecules have small spatial and temporal extent at the source. A
very similar source used to make LiH produces the molecules in the
5~$\mu$s following the firing of the ablation laser. This was
determined from studies with dual ablation pulses having a variable
time separation \cite{LiH}. We believe that the YbF molecules are
formed within a similar temporal window. The initial size of the YbF
pulse is measured by comparing the temporal profile of the molecular
packet at two distances from the source, using the pump beam as one
LIF detector and the probe as the other. Extrapolation back to the
source then gives the initial spatial extent as 11~mm for an initial
5~$\mu$s duration and, in any case, not more than 12~mm.

In demonstrating our field mapping method, we have typically used rf
pulses of 18~$\mu$s duration. During this short time the molecules
move only 11~mm, allowing the rf transition to provide almost a
freeze-frame picture of the molecular pulse at a particular position
inside the interaction region. By studying the centre-frequency,
lineshape and power dependence of the hyperfine transition we can
infer the electric, magnetic and rf fields at that position.
Repeating the experiment for a range of positions allows us to build
a map of the field in the interaction region \footnote{We note that
a somewhat related technique has been used to map the magnetic field
in an atomic fountain clock \cite{NISTfountain}, but this was
limited to the region of the atomic cloud's apogee.}.

By binning the molecules according to arrival time, we can obtain maps with even finer spatial resolution. In the limit where all of the molecules are created at the same time
and in the same place, the arrival time, $t_\mathit{arr}$, is simply
a measure of the molecule's velocity. From this we can infer its
position $z'_\mathit{rf}$ at the time $t_\mathit{rf}$ of the rf
pulse:
\begin{equation*}
z'_\mathit{rf} = \frac{t_\mathit{rf}}{t_\mathit{arr}} l_m\ ,
\end{equation*}
where $l_m$ is the length of the machine from source to detector.
The spatial resolution is limited by three factors: the detector has
an active volume of non-negligible size and hence limited time
resolution, the rf pulses are of finite length, and the beam source
is not point-like in space or time. The uncertainty from the initial
extent of the molecular pulse can be estimated as follows. Assume a
molecule is formed at time $\Delta t$ and position $\Delta l$ with
velocity $v$. The time of arrival at the detector will be
\begin{equation*}
t_\mathit{arr} = \frac{l_m - \Delta l}{v} + \Delta t\ .
\end{equation*}
It is straightforward to show that our estimate of position at time
$t_\mathit{rf}$ will differ from the molecule's true position,
$z_\mathit{rf}$, by
\begin{equation*}
\Delta z_\mathit{rf} = z'_\mathit{rf} - z_\mathit{rf} = \frac{t_\mathit{arr} - t_\mathit{rf}}{t_\mathit{arr}}\ \frac{l_m \Delta t - t_\mathit{arr} \Delta l}{t_\mathit{arr} - \Delta t}\ .
\end{equation*}
Assuming $\Delta t$ and $\Delta l$ to be uncorrelated, with standard
deviations $\sigma_t$ and $\sigma_l$ respectively and with zero
mean, the uncertainty in the estimate of the position is given by
\begin{equation}
\label{source_error}
\sigma_{z_\mathit{rf}} = \frac{t_\mathit{arr} - t_\mathit{rf}}{t_\mathit{arr}^2} \sqrt{t_\mathit{arr}^2 \sigma_l^2  + l_m^2 \sigma_t^2}\ .
\end{equation}
For our source, Eq. \ref{source_error} yields an uncertainty of
11~mm, which is equal to that from the rf pulse duration, giving a
net positional uncertainty of 15~mm (the detector resolution is
small). We note that the effect of finite detector size and rf pulse
duration could  be removed by an appropriate deconvolution, but that
approach is not pursued here.

\section{Field mapping}

\subsection{Electric field measurement}

\begin{figure}[b]
\begin{center}
\includegraphics[width=7.5cm]{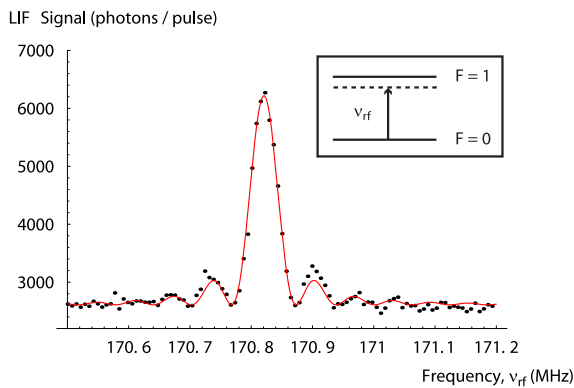}
\caption{Spectrum of the ground-state hyperfine transition.}
\label{rf_lineshape}
\end{center}
\end{figure}

After the $F=1$ state population has been depleted by passing
through the probe laser, it is replenished by the application of an
rf pulse, which transfers population from the $F=0$ state. This
causes an increase in the probe fluorescence, as shown in the rf
frequency scan of Fig. \ref{rf_lineshape}. When considering the
state of the molecule, it is appropriate to chose the static
electric field direction as the axis of quantization because the
electric dipole interaction is by far the strongest coupling to the
external fields. Being a TEM wave, the rf magnetic field between the
plates is accurately perpendicular to the static electric field. As
a result of this polarization, the rf field drives the transitions
$(F=0) \rightarrow (F=1,m_F = \pm1)$. Transitions to $(F=1, m_F=0)$
are strongly suppressed because of the polarisation and are in any
case Stark shifted to a different resonant frequency. For the
resonance shown in Fig. \ref{rf_lineshape}, the rf pulse was applied
when the molecules were near the middle of the interaction region
and the pulse amplitude was chosen to give the molecules a
$\pi$-flip on the Bloch sphere at resonance. The transition
probability peaks at 170.821~MHz, not at the field-free frequency of
170.254~MHz. Because the Stark effect of the YbF hyperfine levels is
accurately known \cite{YbF:LRDR}, this shift allows us to determine
the strength of the electric field at the place where the molecules
are. The Doppler shift is only a few hundred Hertz and is therefore
insignificant at this level of accuracy. The curve shown in Fig.
\ref{rf_lineshape} is the lineshape calculated for a square rf
pulse. This agrees quite well with the data points, though not
perfectly. The deviations, mainly in the wings, are caused by the
departure from a perfectly square rf pulse profile due to the
few-microsecond settling time of the rf switches.

\begin{figure}[b]
\begin{center}
\subfigure[\ Centre of machine]{
\label{rf_2d_1} \includegraphics[width=4cm]{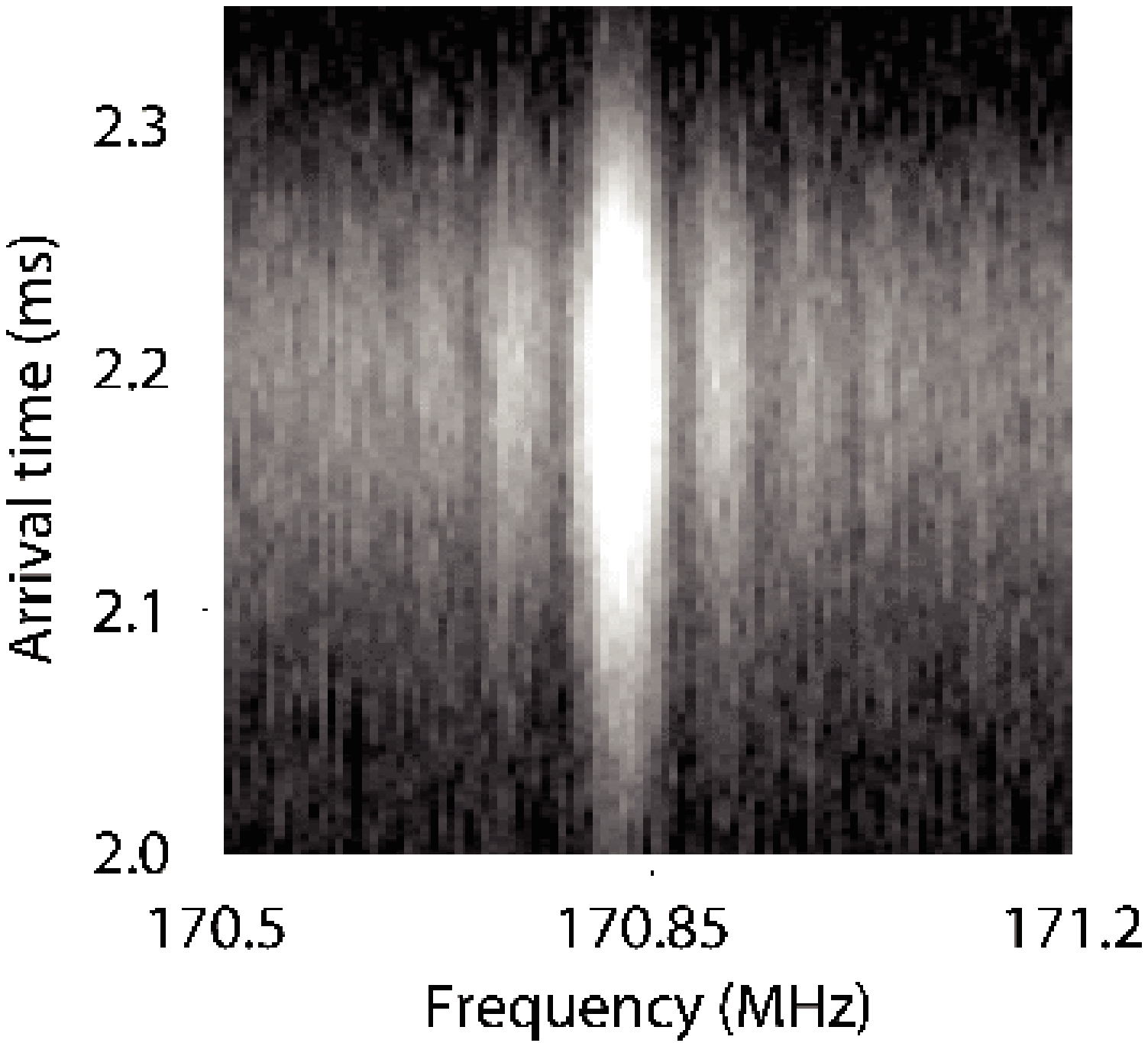}
}
\subfigure[\ Plate end]{
\label{rf_2d_plate_end} \includegraphics[width=4cm]{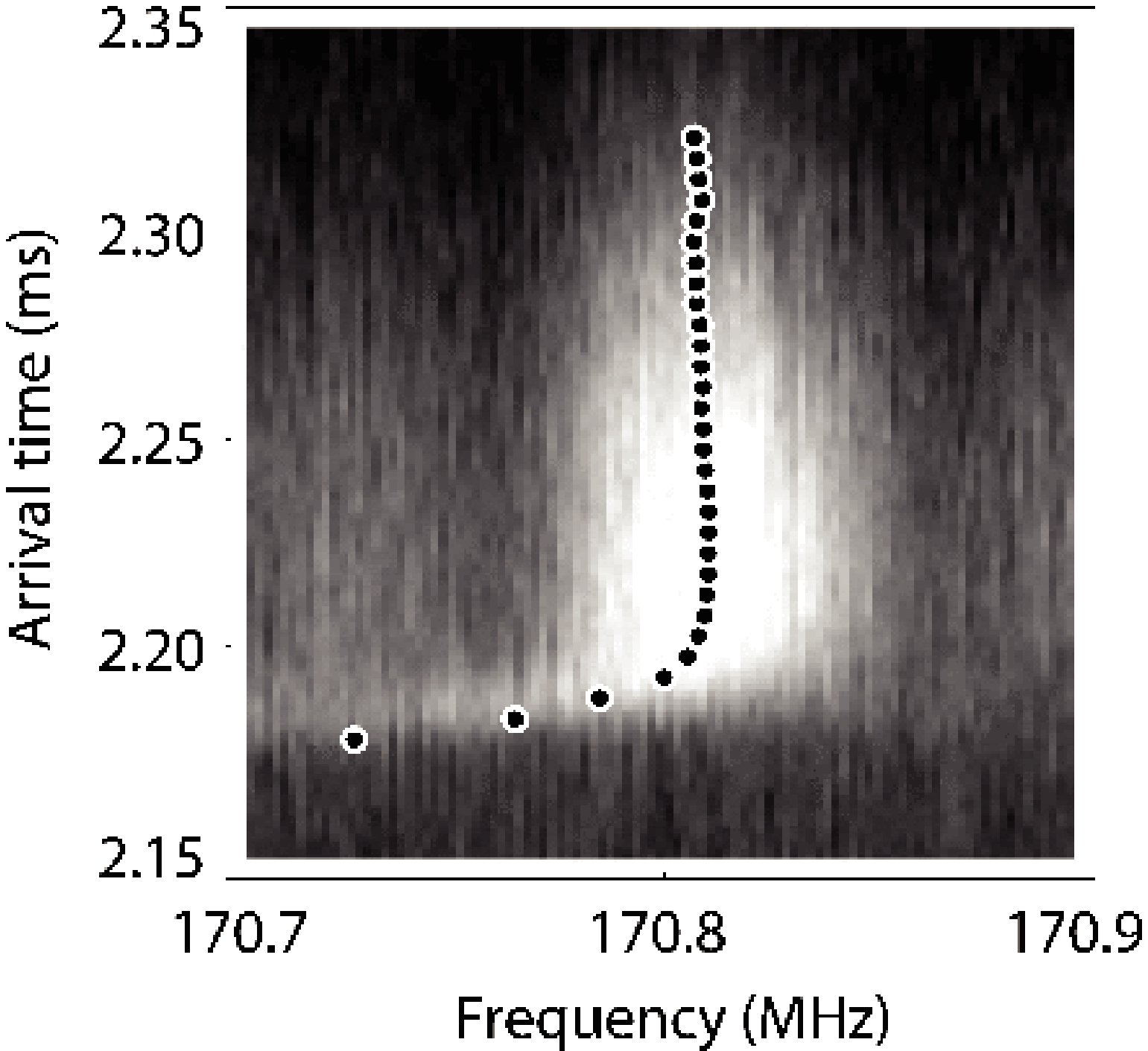}
} \caption{(a) Spectrum of the ground state hyperfine transition as
a function of frequency and arrival time. Integrating this data over
arrival time yields Fig.~\ref{rf_lineshape}. Integrating along the
frequency axis gives a time-of-flight profile, similar to that of
Fig.~\ref{tof}. (b)  Hyperfine transition near the end of the
interaction region. Note that both axes are expanded and shifted
compared to those in (a). Circles show the fitted transition
frequencies for several time slices. }
\end{center}
\end{figure}

\begin{figure*}[htb]
\begin{center}
\includegraphics[width=18cm]{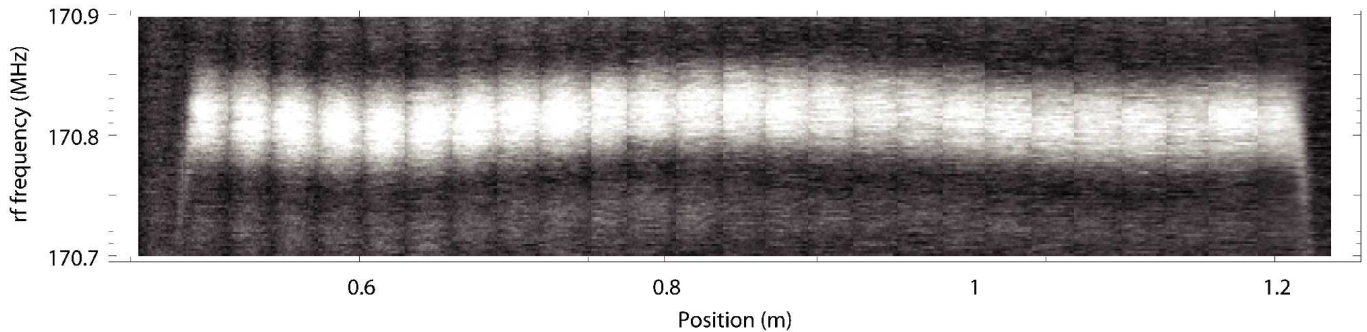}
\caption{Hyperfine transition spectrum as a function of position
along the length of the interaction region. Joins between
overlapping scans can be seen as dark vertical stripes.}
\label{stark_stripe}
\end{center}
\end{figure*}

In figure \ref{rf_lineshape}, the LIF signal is the total photons
counted per beam pulse, i.e. integrated over arrival time. Figure
\ref{rf_2d_1} shows the same data, but here the LIF signal is
plotted versus arrival time as well as frequency \footnote{These and
all subsequent density plots have been gamma-transformed and
windowed to improve the contrast.}. It can be seen that the
lineshape and line-centre are essentially independent of arrival
time, indicating that molecules at the front of the packet (early
arrivals) experience the same static electric field during the rf
pulse as those at the back (late arrivals). This shows that the
electric field is homogeneous in the centre of the interaction
region. In Fig. \ref{rf_2d_plate_end} the rf pulse is applied just
as the molecule leaves the high-voltage plates and the frequency
scan is zoomed in on the central fringe of the resonance. Here we see
that the centre frequency (indicated by the dots) is constant for the slow
molecules, which are still between the plates when the rf is pulsed
on, but shifts dramatically for the fast molecules which explore the region of decreasing field at the exit of the plates.

\begin{figure}[b]
\begin{center}
\includegraphics[width=8cm]{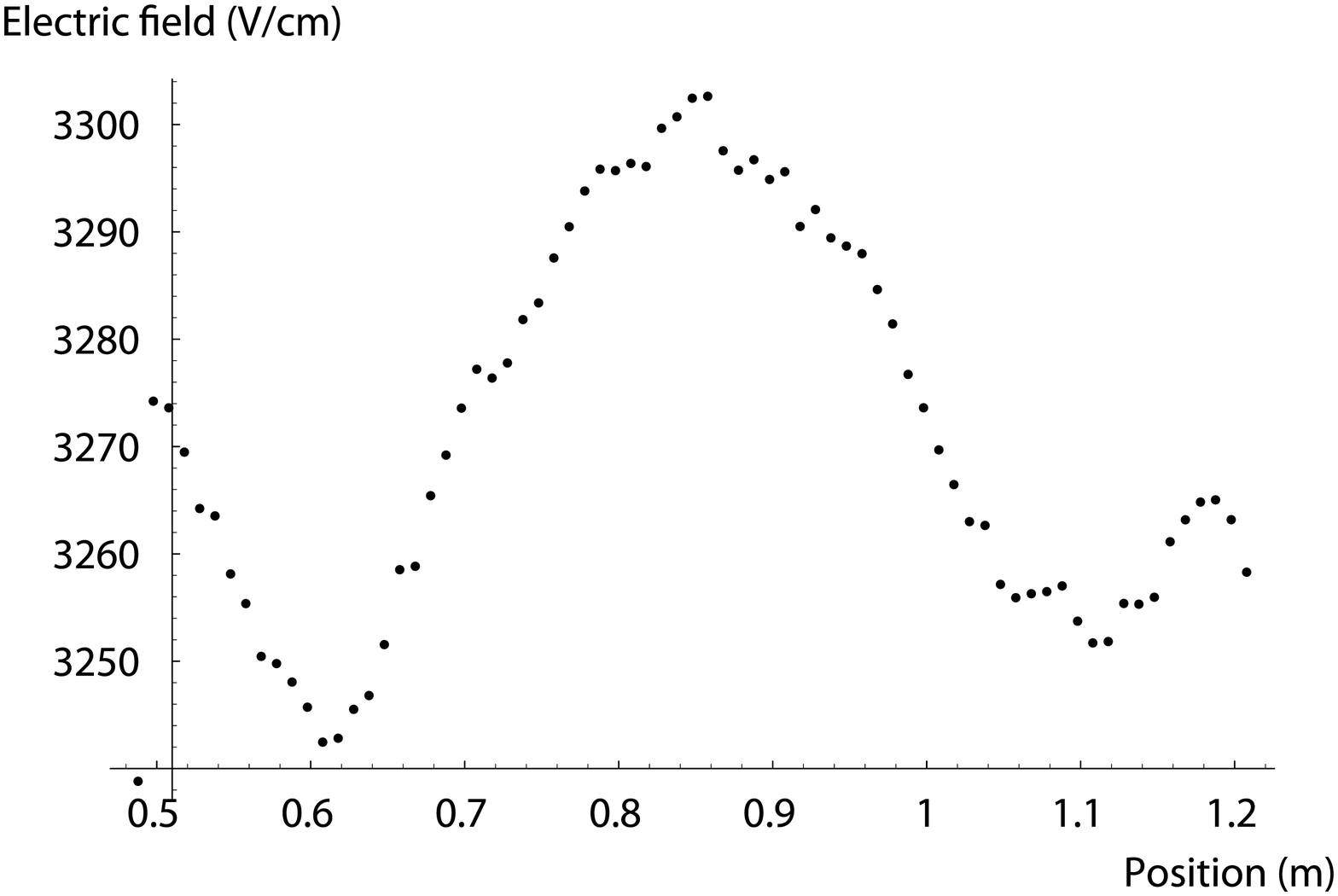}
\caption{Electric field as a function of position. The applied
potential was roughly $\pm$2~kV over the 12~mm gap.} \label{e_map}
\end{center}
\end{figure}

This method of binning the data by arrival time allows us to map out
a region, typically 6~cm in length, with a single frequency scan. In
order to map the electric field over the whole interaction region,
we repeat the scan for several rf pulse delays to make a sequence of
short maps that can be stitched together, as shown in Fig.
\ref{stark_stripe}. This data set was collected in approximately 15
minutes. After fitting the line centre at each position and
converting this to electric field \cite{YbF:LRDR}, we obtain the
field map plotted in Fig. \ref{e_map}. This shows very clearly a
systematic $\pm$110$~\mu$m variation of the plate spacing, which we
can correlate with the design of the support mechanism. The
importance of this measurement is that it shows the variation of
electric field directly and in situ, which mechanical measurements
on the bench cannot do.

The hyperfine transition frequency is sensitive to the electric
field up to very high strengths: for YbF the gradient is 100~Hz per
V/cm at 30~kV/cm. Because of this, polar molecules such as YbF offer
unique sensitivity for probing very high fields. Uncertainty in the
electric dipole moment of the molecule limits the absolute accuracy
to approximately 1\% in the case of YbF \cite{YbF:LRDR}. However,
small relative field measurements are typically limited only by the
accuracy of determining changes in the centre of the line. This in
turn depends on the linewidth associated with the duration of the rf
pulse. In the present experiment, using scans with 15~mm spatial
resolution ($18\,\mu$s pulse), we achieve an uncertainty of 500~Hz
with 30s of measurement, corresponding to a relative field
uncertainty of only 1.5~V/cm at 3.3~kV/cm.

\subsection{Static magnetic field}

If there is a magnetic field component $B_z$ parallel to the
electric field, the $\left(F=1,m_F = \pm1\right)$ energy levels are
split by the Zeeman interaction.  Although this does not shift the
centre of the resonance, it does cause a splitting of the line,
which can be used to measure $B_z$. By contrast, a perpendicular
component of magnetic field has no significant effect on the
lineshape and produces only a very small shift of the centre. This
insensitivity to the perpendicular field is due to the strong tensor
Stark effect, which shifts the $\left(F=1,m_F = 0\right)$ state far
away in energy compared with the off-diagonal magnetic coupling
\cite{EDMPRL}. By switching off the electric field it is also possible
to measure $B_y$, but we do not demonstrate that here.

\begin{figure}[htb]
\begin{center}
\includegraphics[width=7.5cm]{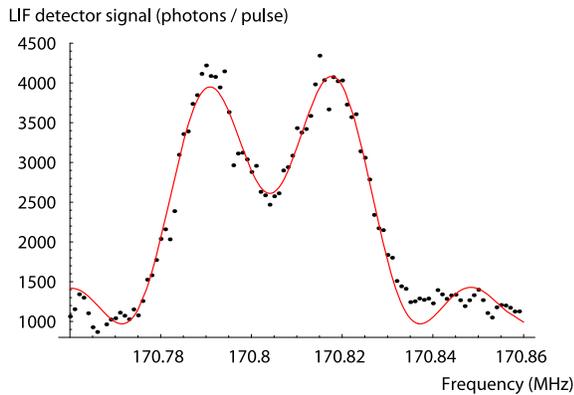}
\caption{Zeeman split hyperfine transition. The fit is to a sum of
two square-pulse transition lineshapes.} \label{zeeman}
\end{center}
\end{figure}

In order to demonstrate magnetic field mapping, we apply a small
static magnetic field, roughly parallel to the strong electric
field, by passing a current $I$ through through copper wires that run parallel to the beam inside
the innermost magnetic shield. In
addition there is a residual ambient field from imperfect shielding,
and perhaps from ferromagnetic impurities within the interaction
region.

Figure \ref{zeeman} shows the resonance line measured near the
centre of the machine with a Zeeman splitting of 25~kHz due to the
application of a $0.9\,\mu$T field. The rf pulse duration was
lengthened to 50~$\mu$s so that this splitting could be clearly
resolved. The theoretical line profile describes the central part of
the line quite adequately, but does not fit well in the wings
because it is once again based on a square rf pulse. In order to
separate the ambient field from the applied field, we vary the
current $I$ and fit the field to the form $B_z=\beta I+B_0$. Once
this is completed at a given position, the rf pulse timing can be
changed and the measurements repeated to build up maps of $\beta$
and $B_0$ along the beamline.

\begin{figure}[htb]
\begin{center}
\subfigure[\ Applied field]{
\label{applied_b} \includegraphics[width=4cm]{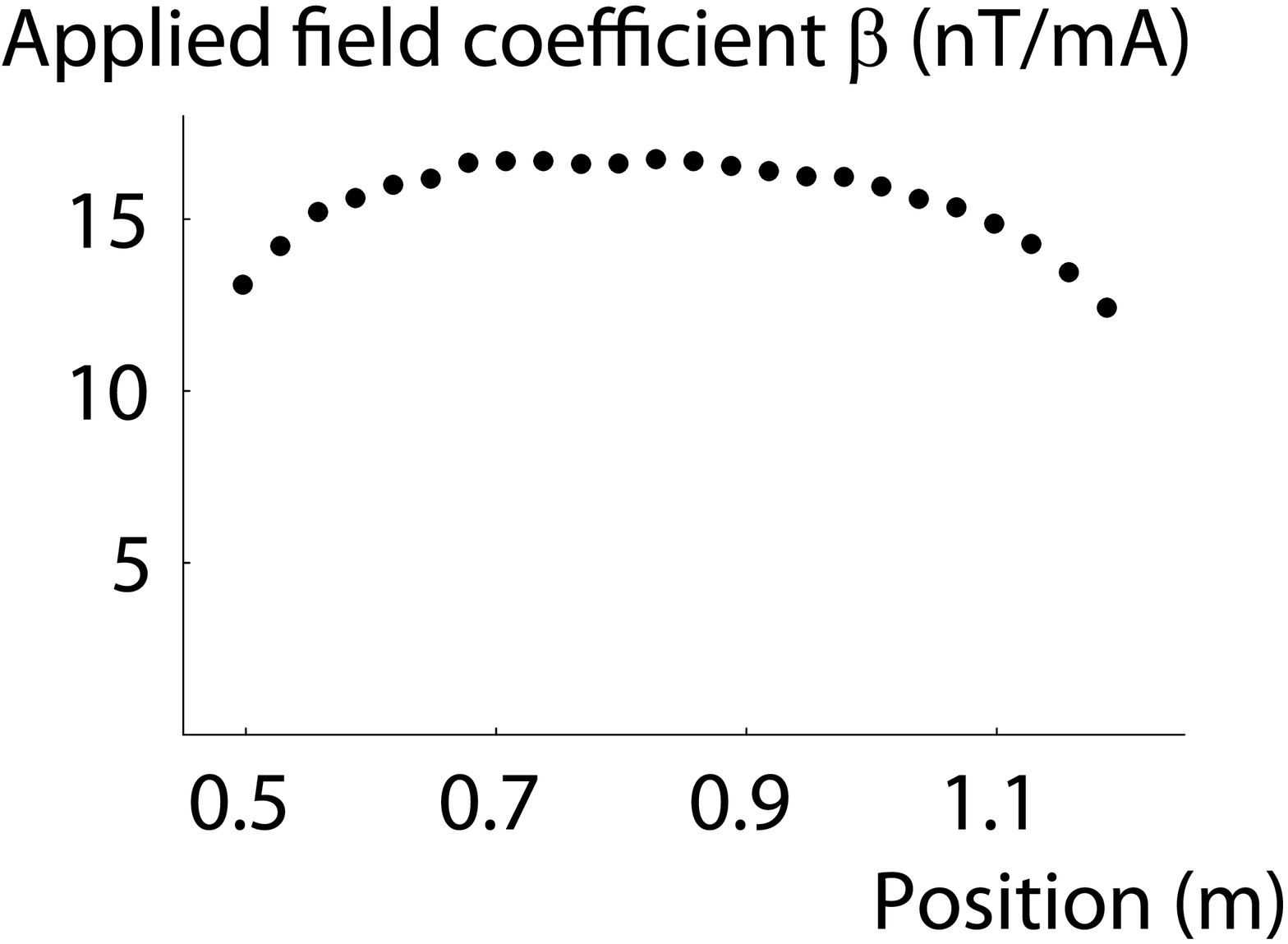}
} \subfigure[\ Ambient field]{
\label{residual_b} \includegraphics[width=4cm]{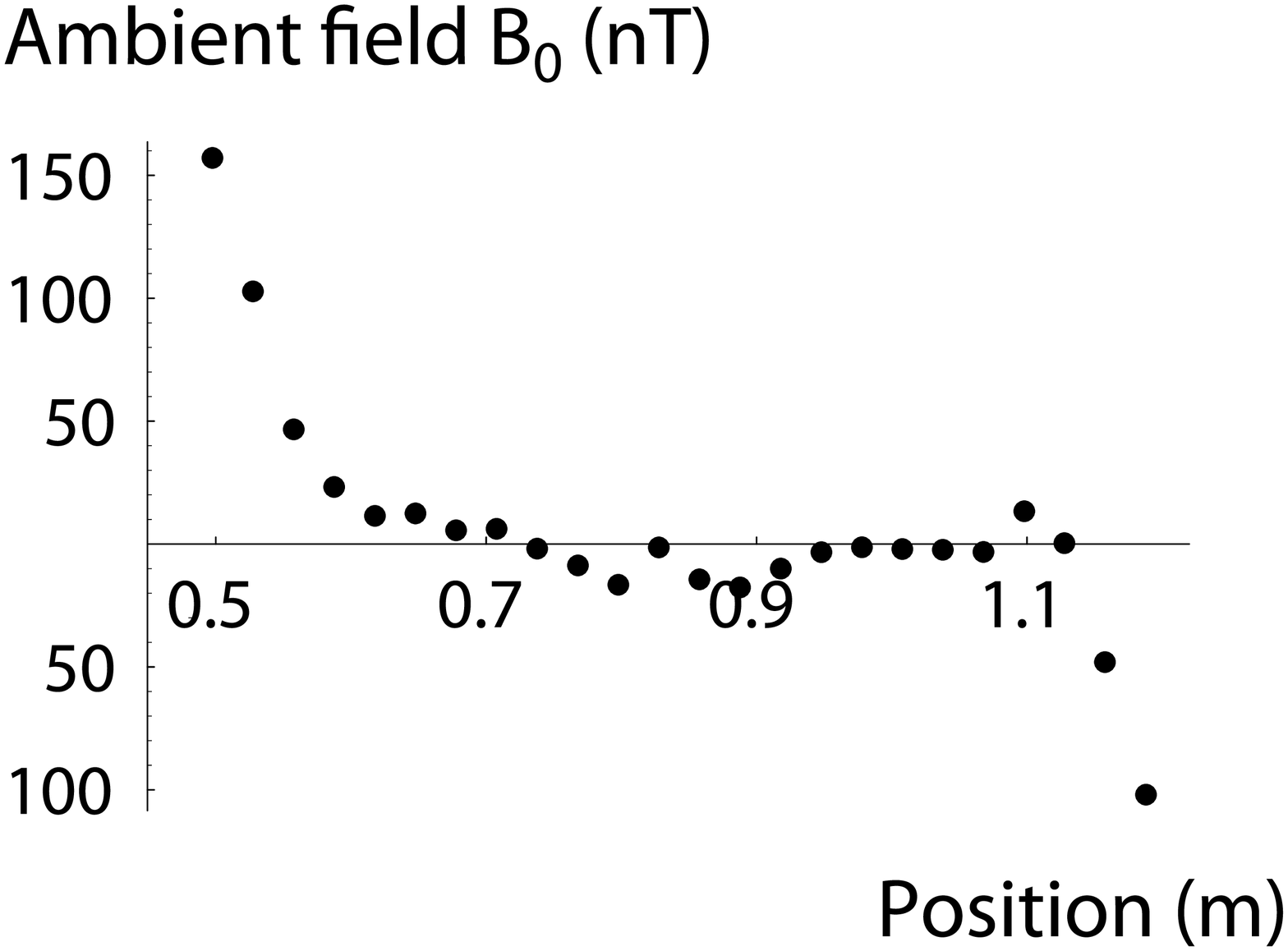}
} \caption{Maps of (a) applied field coefficient $\beta$ and (b)
ambient field $B_0$ in the interaction region.} \label{b_maps}
\end{center}
\end{figure}

Figure \ref{b_maps} shows the results obtained after approximately 4
hours of data collection. The Zeeman splitting is fitted typically
to within 100~Hz. This fitting error leads to an uncertainty in $B_0$ of about 5~nT,
and an uncertainty in $\beta$ of 100~pT/mA. In addition, the
absolute calibration of the field is limited by a 2\% uncertainty in
the g-factor. In Fig.~\ref{applied_b}, $\beta$ decreases at each end
of the interaction region because the field rotates away from the
z-axis to satisfy the boundary condition on the end-caps of the
magnetic shield. The background field shown in Fig.~\ref{residual_b}
exhibits a sharp increase in magnitude at both ends of the
interaction region. This was not expected and requires further
investigation, though we surmise that it is due to inadequate
demagnetisation of the magnetic shield end caps.

\subsection{rf magnetic field}

In the previous two sections, the centre and lineshape of the rf
transition have informed us about the static electric and magnetic
fields. In this section we use the dynamical evolution of the
transition to provide information about the rf magnetic field
itself.

\begin{figure}[htb]
\begin{center}
\includegraphics[width=7.5cm]{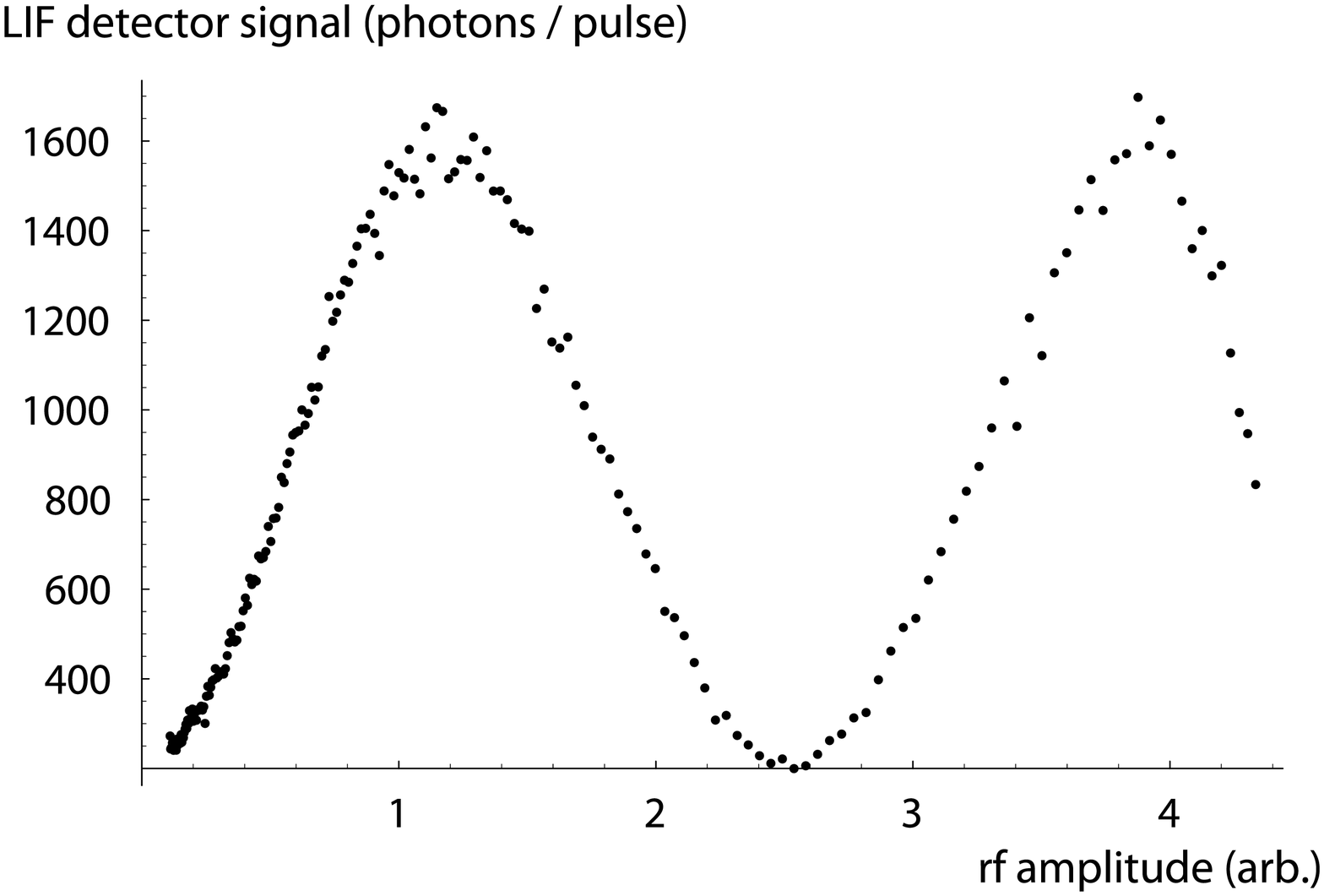}
\caption{Number of molecules in the $F=1$ state versus rf amplitude.} \label{rabi_flop}
\end{center}
\end{figure}

\begin{figure}[b]
\begin{center}
\subfigure[\ Homogeneous field]{ \label{rabi_2d}
\includegraphics[width=4cm]{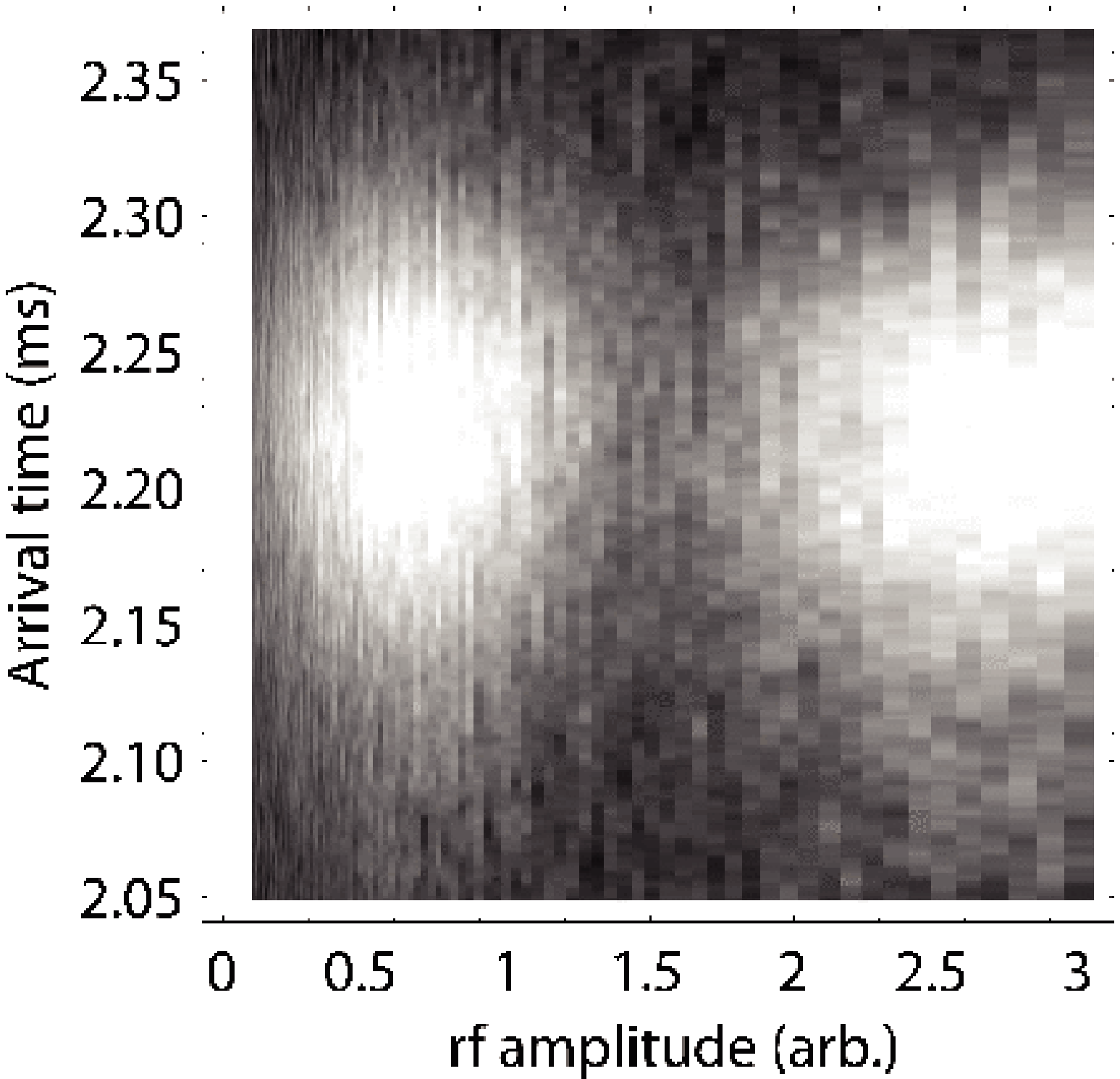} } \subfigure[\
Inhomogeneous field]{ \label{rabi_loop_2d}
\includegraphics[width=4cm]{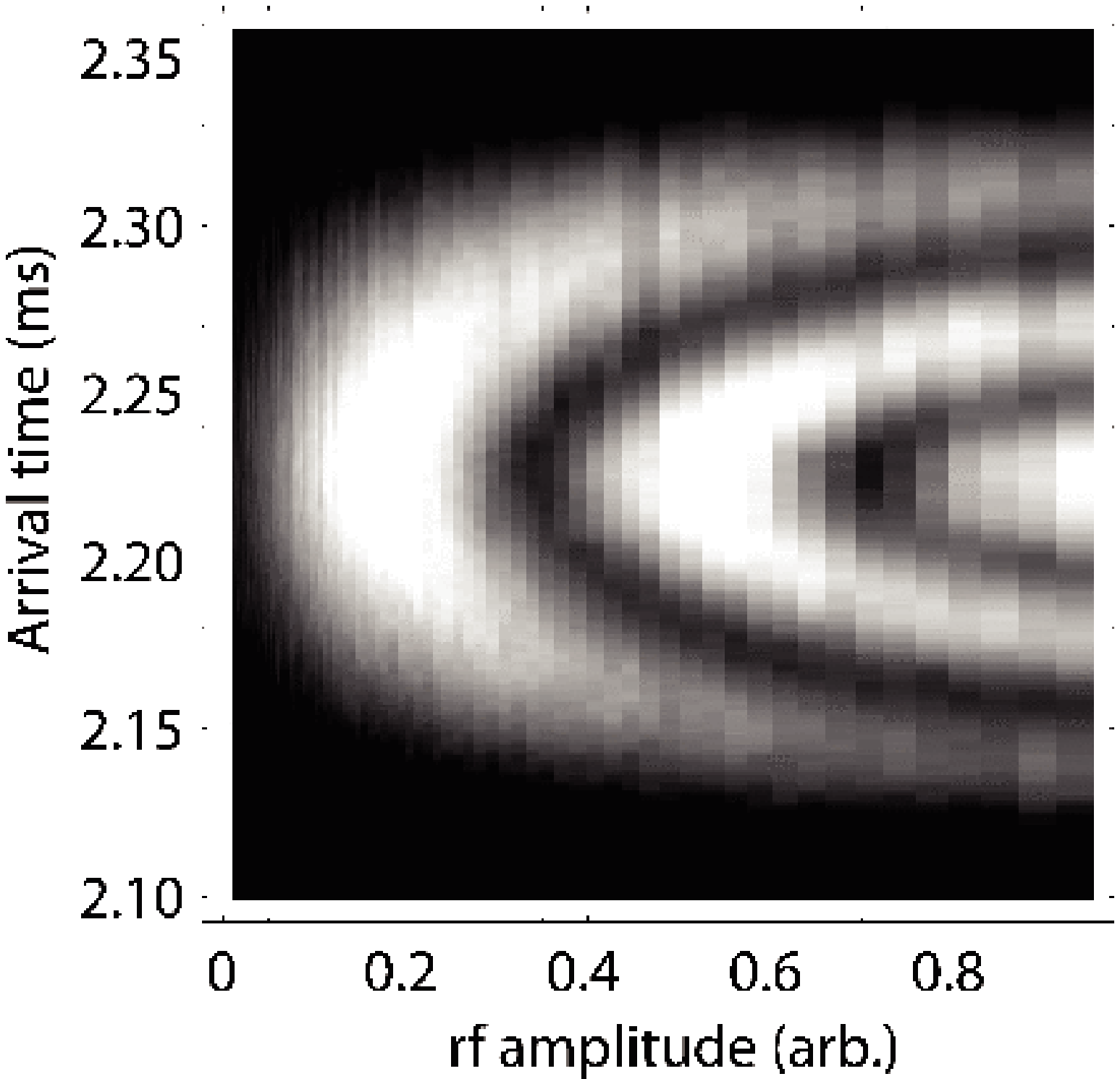} }
\caption{Rabi-flopping, resolved by arrival time: (a) driven by the
homogeneous field of the transmission line, (b) driven by the
inhomogeneous field of an 8~cm diameter loop.}
\end{center}
\end{figure}

The rf field supported by the parallel plate transmission line could
be a pure travelling wave in principle, but imperfect impedance
matching at the ends inevitably produces some standing-wave
component. It is therefore of interest to map the rf
field strength as a function of position along the beamline. For
this measurement, we fix the rf frequency at resonance and scan the
amplitude to observe the Rabi flopping of the molecular population
between the $F=0$ and $F=1$ hyperfine states. Figure \ref{rabi_flop}
shows such a scan covering almost $4\pi$ of Rabi flopping, with the
rf pulse timed to excite molecules at the centre of the interaction
region. The very good visibility of these fringes indicates that the
rf field is rather homogeneous over the beam. In Fig.~\ref{rabi_2d} we explore
the longitudinal direction by replotting the data 
versus arrival time as well as field amplitude, much as we did in
Fig.~\ref{rf_2d_1}. This shows that molecules in different positions
along the length of the packet have the same Rabi flopping rate,
further demonstrating the uniformity of the field.

\begin{figure}[!t]
\begin{center}
\subfigure[\ Loop]{ \label{loop_magnetic_field}
\includegraphics[width=4cm]{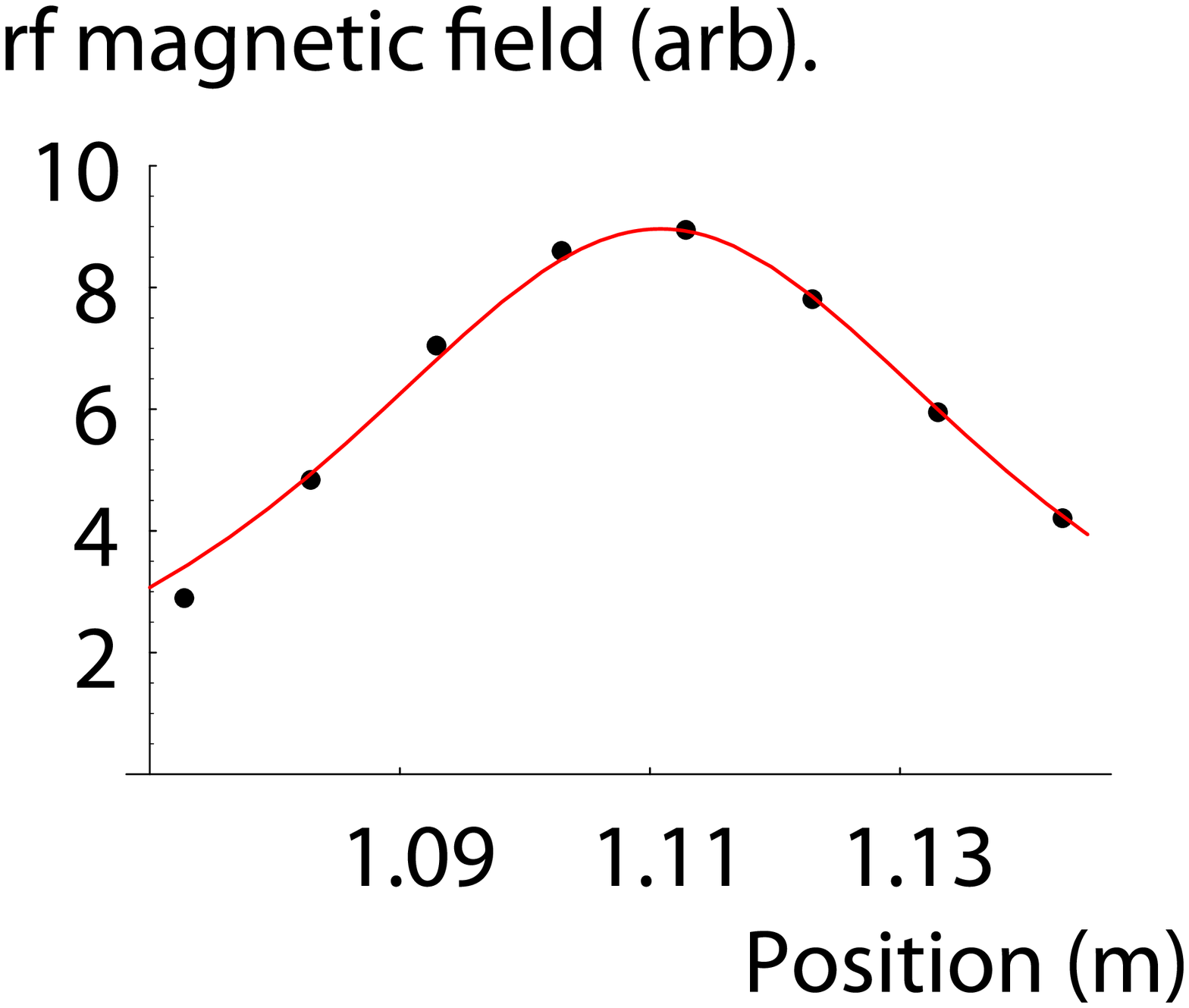}

} \subfigure[\ Transmission line]{ \label{plate_magnetic_field}
\includegraphics[width=4cm]{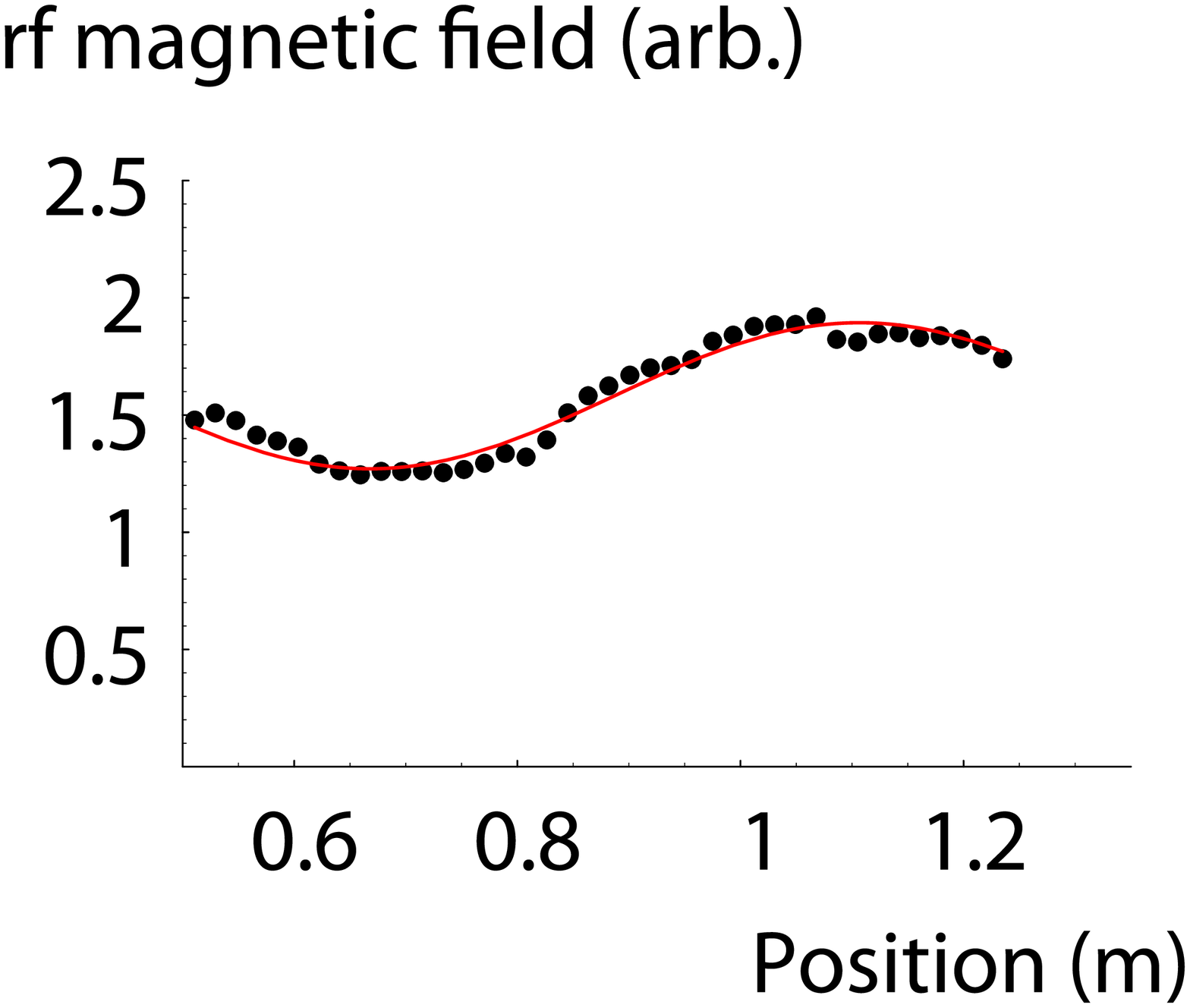} }
\caption{(a) Points: magnetic field of rf loop derived from data of
fig.~\ref{rabi_loop_2d}. Line: fit to dc field of a loop. Peak
strength is the only parameter. (b) Points: rf field in the
transmission line. Line: sine-curve corresponding to the field
amplitude for a 170.820~MHz standing wave. The amplitude and phase
of the standing wave, and amplitude of the travelling wave were
floated in the fit. Note the different x-axis scale compared to
(a).}
\end{center}
\end{figure}

In contrast to this, Fig.~\ref{rabi_loop_2d} shows the result
obtained with an inhomogeneous rf field generated by a copper loop
encircling the field plates. The rf pulse timing was chosen so as to
centre the molecules in the plane of the loop. One sees that
molecules at the centre of the packet, with an arrival time of
approximately 2.24\,ms, undergo a Rabi rotation of $5 \pi$ at the
highest rf amplitude plotted, whereas molecules at the edges of the
packet can barely make a single $\pi$-flip. The variation of the rf
magnetic field amplitude reconstructed from this is shown in
Fig.~\ref{loop_magnetic_field}. The line superimposed on the data is
the simple magnetic field variation expected for a DC field on the
axis of a loop. 

Returning to the transmission line, we have measured the rf field
along the whole length of the interaction region by varying the
timing of the pulse and making many overlapping scans. Figure
\ref{plate_magnetic_field} shows the result and clearly reveals the
standing-wave component of the field. Superimposed on the data is a
sine curve of half the free-space wavelength (87.5~cm), which fits
well to the observed modulation. From this map, we estimate that the
reflected amplitude at each end of the transmission line is 20\%, as
one would expect in view of the impedance step between $50\,\Omega$
and $34\,\Omega$.

\section{Conclusion}
We have described a set of sophisticated diagnostic techniques that
can be applied to pulsed atomic and molecular beams. These
techniques have been developed in our lab to control and measure
systematic effects associated with our electron-edm measurement.
However, the main point of this article is to note the wider
applicability of these methods to any experiment using pulsed atomic
or molecular beams. In particular, these methods are very powerful
in the context of high-precision atomic and molecular physics
experiments, where pulsed beams have not hitherto found widespread
application.


\begin{thebibliography}{[0]}

\bibitem{Scoles} G. Scoles (ed), \emph{Atomic and Molecular Beam Methods} (Oxford University Press,  1988).

\bibitem{Ramsey} N. F. Ramsey, \emph{Molecular beams} (Oxford University Press, 1956).

\bibitem{fountainreview} R. Wynands and S. Weyers, Metrologia \textbf{42}, S64-S79 (2005).

\bibitem{clockreview} J. Vanier and C. Audoin, \emph{The Quantum Physics of Atomic Frequency Standards} (IOP Publishing, 1989).

\bibitem{CsParity} C. S. Wood, S. C. Bennett, D. Cho, B. P. Masterson, J. L. Roberts, C. E. Tanner, and C. E. Wieman, Science \textbf{275}, 1759 (1997).

\bibitem{haensch1s2s} M. Niering et al., Phy. Rev. Lett. \textbf{84}, 5496 (2000).

\bibitem{haenschconstants} M. Fischer et al., Phys. Rev. Lett. \textbf{92}, 230802 (2004).

\bibitem{ohye} Eric R. Hudson, H. J. Lewandowski, Brian C. Sawyer, and Jun Ye, Phys. Rev. Lett. \textbf{96}, 143004 (2006).

\bibitem{nh3} J. van Veldhoven, J. K\"upper1, H.L. Bethlem, B. Sartakov, A.J.A. van Roij, and G. Meijer, Eur. Phys. J. D \textbf{31}, 337 (2004).

\bibitem{Commins94} B. C. Regan, Eugene D. Commins, Christian J. Schmidt, and David DeMille, Phys. Rev. Lett. \textbf{88}, 071805 (2002).

\bibitem{EDMPRL}J. J. Hudson, B. E. Sauer, M. R. Tarbutt and E. A. Hinds, Phys. Rev. Lett. \textbf{89}, 023003 (2002).

\bibitem{fortson} Norval Fortson, Patrick Sandars and Stephen Barr, Phys. Today. \textbf{56}, 33 (2003).

\bibitem{ICAP} B. E. Sauer, H. T. Ashworth, J. J. Hudson, M. R. Tarbutt, and E. A. Hinds, in \emph{Atomic Physics 20}, edited by Christian Roos, Hartmut Haeffner, Rainer Blatt (AIP, New York, 2006),  AIP Conf. Proc. 869,  p. 44.

\bibitem{TarbuttSource} M. R. Tarbutt, J. J. Hudson, B. E. Sauer, E. A. Hinds, V. A. Ryzhov, V. L. Ryabov, and V. F. Ezhov, J. Phys. B \textbf{35}, 5013 (2002).

\bibitem{LiH} S. K. Tokunaga, J. O. Stack, J. J. Hudson, B. E. Sauer, E. A. Hinds, and M. R. Tarbutt, J. Chem. Phys. \textbf{126}, 124314 (2007).

\bibitem{YbF:LRDR} B. E. Sauer, J. Wang and E. A. Hinds, J. Chem. Phys. \textbf{105}, 17 (1996).

\bibitem{NISTfountain} S. R. Jefferts et al., Metrologia \textbf{39}, 321 (2002).

\end{thebibliography}
\end{document}